\documentclass[12pt,preprint]{aastex}

\shorttitle{Gas-grain chemistry in molecular clouds}
\shortauthors{Vasyunin et al.}

\begin{document}

\title{A unified Monte Carlo treatment of gas-grain chemistry for
large reaction networks. I. Testing validity of rate equations in molecular clouds}

\author{A.I. Vasyunin}
\affil{Max Planck Institute for Astronomy, K\"onigstuhl 17, D-69117
Heidelberg, Germany} \email{vasyunin@mpia.de}

\author{D.A. Semenov}
\affil{Max Planck Institute for Astronomy, K\"onigstuhl 17, D-69117
Heidelberg, Germany} \email{semenov@mpia.de}

\author{D.S. Wiebe}
\affil{Institute of Astronomy of the Russian Academy of Sciences,
Pyatnitskaya str. 48, 119017 Moscow, Russia}
\email{dwiebe@inasan.ru}

\and

\author{Th. Henning}
\affil{Max Planck Institute for Astronomy, K\"onigstuhl 17, D-69117
Heidelberg, Germany} \email{henning@mpia.de}

\begin{abstract}
In this study we demonstrate for the first time that the unified Monte Carlo
approach can be applied to model gas-grain chemistry in large reaction networks.
Specifically, we build a time-dependent gas-grain chemical model of the
interstellar medium, involving about 6000 gas-phase and 200 grain
surface reactions. This model is used to test the validity of the standard and
modified rate equation methods in models of dense
and translucent molecular clouds
and to specify under which
conditions the use of the stochastic approach is desirable.

Two cases are considered: (1) the surface mobility of
all species is due to thermal hopping, (2) in addition to thermal
hopping, temperature-independent quantum tunneling for H and H$_2$
is allowed. The physical conditions characteristic for the core and
the outer region of TMC1 cloud are adopted. The gas-phase rate file
RATE\,06 together with an extended set of gas-grain and surface
reactions is utilized.

We found that
at temperatures 25--30~K gas-phase abundances of H$_2$O, NH$_3$, CO
and many other gas-phase and surface species in
the stochastic model differ from those in the deterministic models by
more than an order of magnitude, at least, when
tunneling is accounted for and/or diffusion energies are 3x lower than
the binding energies. In this case, surface reactions, involving light species,
proceed faster than accretion of the same species.
In contrast, in the model without tunneling
and with high binding energies, when the typical timescale of a surface
recombination is greater than the timescale of accretion onto the
grain, we obtain almost perfect agreement between results of
Monte Carlo and deterministic calculations in the same temperature range.
At lower temperatures ($\sim10$~K) gaseous
and, in particular, surface abundances of most important molecules
are not much affected by stochastic processes.
\end{abstract}


\section{Introduction}
Chemical processes are discrete in nature, and it has been realized
long time ago that their microphysically correct theoretical
treatment should rest upon stochastic methods \citep{Gillespie76}.
While in cellular biology Monte Carlo simulations of chemical
processes in cells are widespread, most models in astrochemistry are
based on deterministic rate equation (RE) approach, which has been
proved to be inaccurate for chemical processes on ``cell-like"
objects like dust grains and PAHs
\citep{TielensHagen82,Caselli_ea98,
hershem03,LipshtatBiham03,StantchevaHerbst04}. The concept of
mass-action kinetics, which lies behind rate equations, being
adequate for gas-phase reactions, is not appropriate in the case of
small surface population (number of particles $\leq1$ per
grain). This situation is often the case in grain surface catalysis.
Rates of surface reactions obtained through this approach may be
strongly incorrect, leading to an improper estimation of timescales
of fundamental processes, like H$_{2}$ formation, and significant
errors in abundances of other species.

Up to now, a number of attempts have been made to assess the
significance of stochastic effects for astrochemical simulations and
to develop suitable numerical methods.
In the context of astrochemistry first attempt to treat grain
surface chemistry stochastically has been made by
\cite{TielensHagen82}. In this study, equilibrium abundances of
gas-phase species were used to obtain accretion rates and to
calculate time-dependent populations of surface species through
Monte Carlo approach. All surface species but H$_{2}$ were assumed
to remain on the grain surfaces.

Subsequent attempts to account for the discrete nature of
surface reactions can be divided into two categories. The first
category encompasses different modifications to the rate equations
for grain surface chemistry \citep{Caselli_ea98, Caselli_ea02,
Stantcheva_ea01}. The main idea behind these modifications is to
restrict rates of diffusive surface reactions by accretion or
desorption rates of reactants. Thus, the so--called
accretion-limited regime is taken into account, when the rate of a
particular surface reaction is determined by the flux of reactants
accreting on a grain and not by their diffusion over the surface.
For simple reacting systems, results obtained with this
semi-empirical approach were found, at least, to be in closer
agreement with Monte Carlo (MC) computations than the treatment by
rate equations alone. However, the applicability of the modified
rate equations (MRE) to large chemical networks has not been tested
so far. Moreover, it is impossible to investigate purely stochastic
effects, like bistability, with this technique, still purely
deterministic in its nature \citep{LeBourlot_ea93,
ShalabieaGreenberg95, Lee_ea98, BogerSternberg06}.

In the second category of studies the master equation is solved with
different methods. Because the chemical master equation describes
probabilities of a chemically reacting system to be in all possible
states of its phase space and the number of these states grows
exponentially with the number of species, their direct integration
is only possible for extremely simple systems, consisting of a very
few different components.
\cite{Biham_ea01}, \cite{Green_ea01} and \cite{Lipshtat_ea04} have
performed a direct integration of the master equation to investigate
H$_{2}$ formation. These studies demonstrated that in some cases
this fundamental process cannot be adequately described with the
rate equation approach. In \cite{Stantcheva_ea02},
\cite{Caselli_ea02} and \cite{StantchevaHerbst03} direct integration
of the master equation was applied to study the evolution of a
simple H---O---CO surface chemical network and deuterium
fractionation. Direct integration of the master equation in more
complex networks of grain surface reactions is hampered by the lack
of appropriate computing power. To the best of our knowledge, the
only successful simulation of a more extended surface chemical
network in astrochemistry, consisting of 19 reactions, with direct
integration of the master equation was accomplished by
\cite{StantchevaHerbst04}.

A Monte Carlo approach based on continuous-time random-walk method
has been developed by \cite{Chang_ea05} to model recombination of
hydrogen atoms on interstellar grains. The advantage of this method is
that it allows to model formation of H$_2$ and other molecules on surfaces of arbitrary
roughness, which has later been demonstrated by \cite{CuppenHerbst07}.

To relax computational requirements needed to solve the master
equation, the moment equation approximation was suggested by
\cite{LipshtatBiham03} and \cite{BarzelBiham07}. Currently, this is
the most promising approach to efficient simulations of gas-grain
systems with complex grain surface chemistry, because the moment
equations can be easily combined with rate equations used to
simulate gas-phase chemical processes. But, strictly speaking, the
moment equations are an approximation and their validity should be
verified by a comparison with exact methods.

The only feasible technique allowing us to obtain an exact solution
to the master equation for complex chemically reacting systems is
Monte Carlo method developed by \cite{Gillespie76}. This approach is
widely used in molecular biology for simulation of chemical
processes in living cells and can also be applied to astrochemical
problems. Gillespie's Stochastic Simulation Algorithm (SSA) was
first used in complex astrochemical networks by \cite{Charnley98},
where SSA was applied to gas-phase chemistry only. Later a simple
grain surface H---O---CO chemical model, similar to the one
considered in \cite{Caselli_ea98}, was developed by
\cite{Charnley01}. In this model, gas-phase abundances of species
were assumed to be constant and desorption was neglected. No coupled
gas-grain chemical model was studied.

The chemical master equation is usually
considered as a method to model surface reactions. However, due to
its universality, it can be used to model the gas-phase chemistry as
well. The reason why the rate equations are so popular as a tool to
study gas-phase chemistry in astronomical environments is the fact
that they are easier to implement and much more computationally
effective. However, if one uses the rate equations to model
gas-phase chemistry and some stochastic approach to model surface
chemistry one has to find a way to match the two very different
computational techniques. The first study in which a Monte Carlo treatment of grain surface
chemistry was coupled to time-dependent gas-phase chemistry is
\cite{Chang_ea07}. In this paper iterative technique is utilized to
combine very different Monte Carlo and rate equations approaches.
Calculations are performed for the time interval of $2\cdot10^{5}$
years and a relatively simple H---O---CO surface network consisting
of 12 reactions is used.

Even though remarkable progress has been made to develop stochastic
approaches to astrochemistry, up to now there are no studies in
which a stochastic treatment of a complex grain surface chemical
network is fully coupled to time-dependent gas-phase chemistry.
The situation gets less complicated when
the same method is used for both gas-phase and surface chemistry. In
this paper, for the first time, we present a ``complete''
time-dependent chemical gas-grain model, calculated with Monte Carlo
approach. We solve the master equation with a Monte Carlo
technique, using it as a single method to treat gas-phase reactions,
surface reactions, and gas-grain exchange processes simultaneously.
The gas-phase chemical network includes more than 6000 reactions while
surface network consists of more than 200 reactions.This Monte Carlo model is used to test the validity of rate equations and modified
rate equations for a range of physical conditions, typical of
diffuse and dense clouds.
So far, no comparison of RE and MRE techniques against rigorous
stochastic approach over a wide range of
physical conditions, typical for astrophysical objects, has been
made.

There is clear observational evidence that interstellar grains have
complex mineral composition and may have either amorphous or
crystalline structure. In general, it is believed that interstellar
dust particles are made of olivine-like silicates and some form of
amorphous carbon or even graphite \citep{DraineLee78}. The location
on a surface which may be occupied by an accreted molecule is
commonly referred to as a surface site. The density of sites and
binding energies for specific species strongly depend on the dust
material and its structure. Binding energies $E_{\rm b}$, that is,
potential barriers between adjacent sites, have been assumed by
\cite{HHL92} to be roughly 0.3 of corresponding desorption energies
$E_{\rm D}$.
In \cite{Pirronello_ea97a, Pirronello_ea97b, Pirronello_ea99},
\cite{Katz_ea99} and \cite{Perets_ea05} experimental results on the
H$_{2}$ formation, the ratio of diffusion energy to desorption
energy, $E_{\rm b}/E_{\rm D}$, was found to be close to 0.77 for H
atoms. This ratio determines the surface mobility of species and is
of utter importance for estimating surface reaction rates.

Another factor defining rates of reactions involving H and H$_2$
(and their isotopologues) is related to the possibility of their
tunneling. Non-thermal quantum tunneling of these species allows
them to scan grain surface and find a reacting partner quickly. This
mechanism has been suggested as an explanation of efficient H$_2$
formation in the ISM. Later analysis of the cited experimental
results has lead \cite{Katz_ea99} to the conclusion that tunneling
does not happen. The absence of tunneling implies lower mobility, so
that surface processes are essentially reduced to accretion and
desorption. However, this conclusion turned out to be not the final
one. In other studies \citep[e.g.][]{CazauxTielens04} alternative
explanations of these experiments have been proposed and some
shortcomings of the theoretical analysis performed in
\cite{Katz_ea99} have been found. So, up to now the question of
presence or absence of tunneling effects on grain surfaces is not
settled, and the $E_{\rm b}/E_{\rm D}$ ratio is not well
constrained.
With this in mind, two different
models of grain surface reactions are considered: (1) the surface
mobility of all species is caused by thermal hopping only with high $E_{\rm b}/E_{\rm D}$ ratio equal to 0.77, (2)
temperature-independent quantum tunneling for H and H$_2$ is
included in addition to thermal hopping with low $E_{\rm b}/E_{\rm D}$ ratio equal to 0.3.

The organization of the paper is as follows. In Section~2 we
describe physical conditions adopted in simulations and chemical
model used. Section~3 contains basics of stochastic reaction
kinetics and detailed description of the Monte Carlo code used for
simulations. In Section~4 the validity of rate equations and
modified rate equations is checked against Monte Carlo method.
First, global agreement between methods is investigated. Then, some
interesting species are discussed separately. In Section~5 a
discussion is presented. Section~6 contains the conclusions.

\section{Modeling}

\subsection{Physical conditions}
\label{sec:cloud_models} In the present study, we consider grain
surface reactions under physical conditions typical of irradiated
translucent clouds, cold dark cores and infrared dark clouds. We consider temperatures $T$
between 10 and 50~K, densities $n(\rm H)$ between 200 and
$2\cdot10^4$~cm$^{-3}$ (three values of $n(\rm H)$ and five values of $T$), and visual extinctions of 0.2~mag at $n({\rm
H})=200$~cm$^{-3}$, 2.0~mag at $n({\rm H})=2\cdot10^3$~cm$^{-3}$ and
15~mag at $n({\rm H})=2\cdot10^4$~cm$^{-3}$
\citep[e.g.,][]{SnowMcCall06,Hassel_ea08}. This $A_{\rm v}$ roughly corresponds
to the distance of order of 0.4--0.5 pc from the cloud boundary (at constant
density). The cloud is illuminated by the mean interstellar diffuse UV field. The dust temperature is
assumed to be equal to gas temperature. We do not study the earliest
stage of the molecular cloud evolution, which is essentially the
stage when atomic hydrogen is converted into hydrogen molecules. The
chemical evolution is simulated assuming static physical conditions.
While it is still a matter of debate how long does it take for a
typical isolated cloud core to become gravitationally unstable and
to start collapsing, we assume a short timescale of 1~Myr which
seems to be appropriate for the TMC1 cloud \citep[see
e.g.][]{Roberts_ea04, Semenov_ea04}.

\subsection{Chemical model}
The chemical model is the same as described in
\citet{Vasyunin_ea08}. The gas-phase reactions and their rates are
taken from the RATE\,06 database, in which the effects of
dipole-enhanced ion-neutral rates are taken into account
\citep{Woodall_ea07}. All reactions with large negative activation
barriers are excluded. The rates of photodissociation and
photoionization of molecular species by interstellar UV photons are
taken from \citet{vDea_06}. The self- and mutual shielding of CO and
H$_2$ against UV photodissociation are computed as described in
\citet{van_Zadelhoff_ea03} using pre-calculated factors from
Tables~10 and 11 from \citet[][]{Lee_ea96}. Ionization and
dissociation by cosmic ray particles are also considered, with a
cosmic-ray ionization rate of $1.3\cdot10^{-17}$~s$^{-1}$
\citep{SpitzerTomasko68}.

The gas-grain interactions include accretion of neutral species onto
dust grains, thermal and photodesorption of mantle materials,
dissociative recombination of ions on charged grains and grain
re-charging processes. The dust grains are uniform $0.1\,\mu$m
spherical particles made of amorphous silicates of olivine
stoichiometry \citep{Semenov_ea03}, with a dust-to-gas mass ratio of
$1\%$. The sticking probability is 100\%. Desorption energies
$E_{\rm D}$ and surface reaction list are taken from
\citet{GarrodHerbst06}.

The grain surface is assumed to be compact, with surface density of
$2\cdot10^{14}$~sites\,cm$^{-2}$, which gives $\approx 3\cdot10^5$
sites per grain \citep{Biham_ea01}. We employ two models to
calculate the rates of surface reactions
(Table~\ref{tab:surf_mods}). In Model~T (T for tunneling) the tunneling timescale for
a light atom to overcome the potential barrier and migrate to
another potential well is computed using Eq.~(10) from
\citet{HHL92}, with the barrier thickness of 1\AA. In Model H (H for hopping) we do
not allow H and H$_2$ to scan surface sites by tunneling. The
diffusion timescale for a molecule is calculated as the timescale of
thermal hopping multiplied by the total number of surface site and
is given by Eq.~(2) and Eq. (4) from \citet{HHL92}. The hopping rates are
sensitive to the adopted values of diffusion energy, which are not
well constrained. Thus, we consider low and high diffusion energies
which are calculated from adopted desorption energies by multiplying
them by factors of 0.3 \citep[like in][]{HHL92} (Model T) and 0.77
\citep[][]{Katz_ea99} (Model H), respectively. For all considered
models, the total rate of a surface reaction is calculated as a sum
of diffusion or tunneling rates divided by the grain number density
and multiplied by the probability of reactions (100\% for processes without activation energy).

Overall, our network consists of 422 gas-phase and 157 surface
species made of 13 elements, and 6002 reactions including 216
surface reactions. As initial abundances, we utilize the ``low
metallicity'' set of \citet{Lee_ea98}, where abundances of heavy
elements in the gas are assumed to be severely depleted. All
hydrogen is molecular initially. The chemical evolution for 1~Myr in
the classical deterministic approaches is computed with the fast
``ALCHEMIC'' code\footnote{Available upon request: semenov@mpia.de}
in which the modified rate equations are implemented according to
\cite{Caselli_ea98}. No further modification for reactions with activation energy barriers is used \citep[see][]{Caselli_ea02}.

\section{Stochastic reaction kinetics}

\subsection{Theoretical foundations}
We consider a chemically reacting system which consists of $N$
different types of species \{$S_1$ ... $S_N$\} and $M$ chemical
reactions \{$R_1$ ... $R_M$\} \citep{Gillespie76}. All these species
are contained in a constant volume $\Omega$, in which local thermal
equilibrium is reached, so that the system is well mixed. We denote
the number of species of type $i$ at time $t$ as $X_i(t)$. The
ultimate goal is to determine the state vector $\vec{X}(t)=\{X_1(t)
... X_N(t)\}$ of the system at any given time $t>t_0$, assuming
certain initial conditions $\vec{X}(t_0)=\vec{X}_0$.

Let us assume that each reaction $R_j$ retains properties of a
Markov chain and can be considered as a set of independent
instantaneous events. Each chemical reaction is described by two
quantities, the discrete state vector $\vec{\nu}=\{\nu_{1j}, ...
,\nu_{Nj}\}$ and the propensity function $a_j$. The components of
the state vector $\nu_{ij}$ represent the net change in populations
of the $i$th species due to the $j$th reaction. By definition, the
propensity function
\begin{equation}\label{rrate}
a_j(\vec{X})dt
\end{equation}
is the probability for a reaction $R_j$ to occur in the volume
$\Omega$ over the time interval $dt$. In analogy to the reaction
rate in the rate equation approach, for a one-body
reaction the propensity function $a_j$ can be expressed through rate
constants and numbers of reactants as
\begin{equation}
a_j(\vec{X})=c_jx_1.
\end{equation}
Here, $c_j$ is the rate constant of the $j$th reaction and $x_1$ is
the absolute population of reagent. For a two-body
heterogeneous reaction this expression changes as follows:
\begin{equation}
a_j(\vec{X})=c_jx_1x_2,
\end{equation}
where $c_j$ is the reaction rate constant, $x_1$ and $x_2$ are the
abundances of first and second reactants. If we deal with a two-body
homogeneous reaction (like H~+~H~$\rightarrow$~H$_{2}$), the
expression for its propensity function becomes somewhat more
complicated:
\begin{equation}
a_j(\vec{X})=\frac{c_j}{2}x_1(x_1-1)
\end{equation}
This expression reflects the fact that the rate of a homogeneous
two-body reaction is proportional to the number of all possible
pairs of its reactants. The term ($x_1-1$) is important for a proper
calculation of the reaction rate in the stochastic regime when the
average reagent population is close to unity and cannot be properly
reproduced by the rate equation approach. Note that the abundances
of species are always integer numbers in these expressions.

These definitions allow to characterize the microphysical nature of
chemical processes and to establish a basis for stochastic chemical
kinetics. A vector equation that describes the temporal evolution of
a chemically reacting system by stochastic chemical kinetics is the
chemical master equation. To derive this equation one has to
introduce the probability of a system to be in the state
$\vec{X}(t)$ at a time $t>t_0$:
\begin{equation}\label{pdef}
P(\vec{X},t\mid\vec{X}_0,t_0)\equiv
Prob(\vec{X}(t)=\vec{X}\mid\vec{X}(t_0)=\vec{X}_0).
\end{equation}
Here $P(\vec{X},t\mid\vec{X}_0,t_0)$ is the conditional probability
density function of the time-dependent value $\vec{X}$. Temporal
changes of $\vec{X}$ are caused by chemical reactions with rates
defined by Eq.~(\ref{rrate}). Therefore, the change of
$P(\vec{X},t\mid\vec{X}_0,t_0)$ over the time interval $dt$ is the
sum of probabilities of all possible transitions from the state
$\vec{X_0}$ to the state $\vec{X}(t)$:
\begin{eqnarray*}
P\left(\vec{X},t+dt\mid\vec{X}_0,t_0\right)= \\
P\left(\vec{X},t\mid\vec{X}_0,t_0\right)\cdot\left\lbrack1-
\sum_{j=1}^M\left(a_j(\vec{X})dt\right)\right\rbrack+
\\
\sum_{j=1}^M P(\vec{X}-\vec{\nu_j},t\mid\vec{X}_0,t_0)\cdot
\left(a_j(\vec{X}-\vec{\nu}_j)dt\right)
\end{eqnarray*}
The first term in this equation is derived from the fact that the
system is already in the state $\vec{X}$ and describes the
probability for the system to leave the state during the time
interval $dt$. The second term defines the total probability for the
system to reach the state $\vec{X}$ from states
$\vec{X}-\vec{\nu_j}$ during the time interval $dt$ due to a
reaction $R_j$. Finally, we formulate the chemical master equation:
\begin{eqnarray}\label{mastereq}
\frac{\partial P(\vec{X},t\mid\vec{X_0},t_0)}{\partial t}=
\sum_{j=1}^M\lbrack
a_j(\vec{X}-\vec{\nu_j})P(\vec{X}-\vec{\nu_j},t\mid\vec{X_0},t_0)-
a_j(\vec{X})P(\vec{X},t\mid\vec{X_0},t_0)\rbrack
\end{eqnarray}
This vector equation looks similar to the conventional set of
balance equations. However, it properly takes into account the
discrete nature of chemical processes and, thus, has a much wider
range of applicability. Unfortunately, an analytic integration of
the chemical master equation is only possible in a very limited
number of cases \citep[several species and reactions, see
e.g.][]{StantchevaHerbst04}. Thus, one has to rely on various
numerical techniques to solve this equation.

\subsection{Implementation of the Monte Carlo algorithm}\label{phys_model}

In principle, any normalization of abundances can be used in
Eq.~(\ref{mastereq}). In particular, if we take $\Omega$ to be a
unit volume, we end up with the usual number densities, which are
widely used in astrochemical models. When dealing with mixed
gas-dust chemistry, one has to be more careful with the
normalization. As surface reactions can only proceed when both
reactants reside on a surface of the same grain, we define $\Omega$
as the volume of the interstellar medium that contains exactly one
dust grain,
\begin{equation}\label{gasvol_pergrain}
\Omega_{\rm MC} = \frac{\frac{4}{3}\pi\rho_{\rm
dust}r^{3}}{\rho_{\rm gas}\gamma},
\end{equation}
where $r$ is the grain radius, $\rho_{\rm dust}$ is the mass density
of grain material, $\rho_{\rm gas}$ is the mass gas density, and
$\gamma$ is the dust-to-gas (0.01) mass ratio. Here grains are
assumed to be spheres of equal size. As we also assume that the
interstellar medium is well-mixed, so that the volume $\Omega_{\rm
MC}$ is representative for any volume of the real interstellar
medium with the same physical conditions.

This setup allows to construct the unified master equation for all
three kinds of processes listed above (gas phase reactions,
accretion/desorption processes and grain surface reactions). This
equation is solved with the stochastic Monte Carlo algorithm
described in \citet{Gillespie76}. We implemented this technique in a
FORTRAN77 code which allows to simulate all chemical reactions in
the network in a self-consistent manner. Such a ``brute-force''
approach requires substantial CPU power and cannot be utilized in
massive calculations. In the present study, it is used as a
benchmark method to simulate chemical evolution in astrophysical
objects. A typical run on a single Xeon 3.0GHz CPU takes between 10
hours and several days of computational time, and involves several
billions of time steps. If one is only interested in a single point then
the model can be used not only for benchmarking, but
also for practical purposes. Unfortunately, it becomes impractical
if abundances are computed for a number of spatial locations. High density also
slows down the computation significantly.

\begin{figure}
\plotone{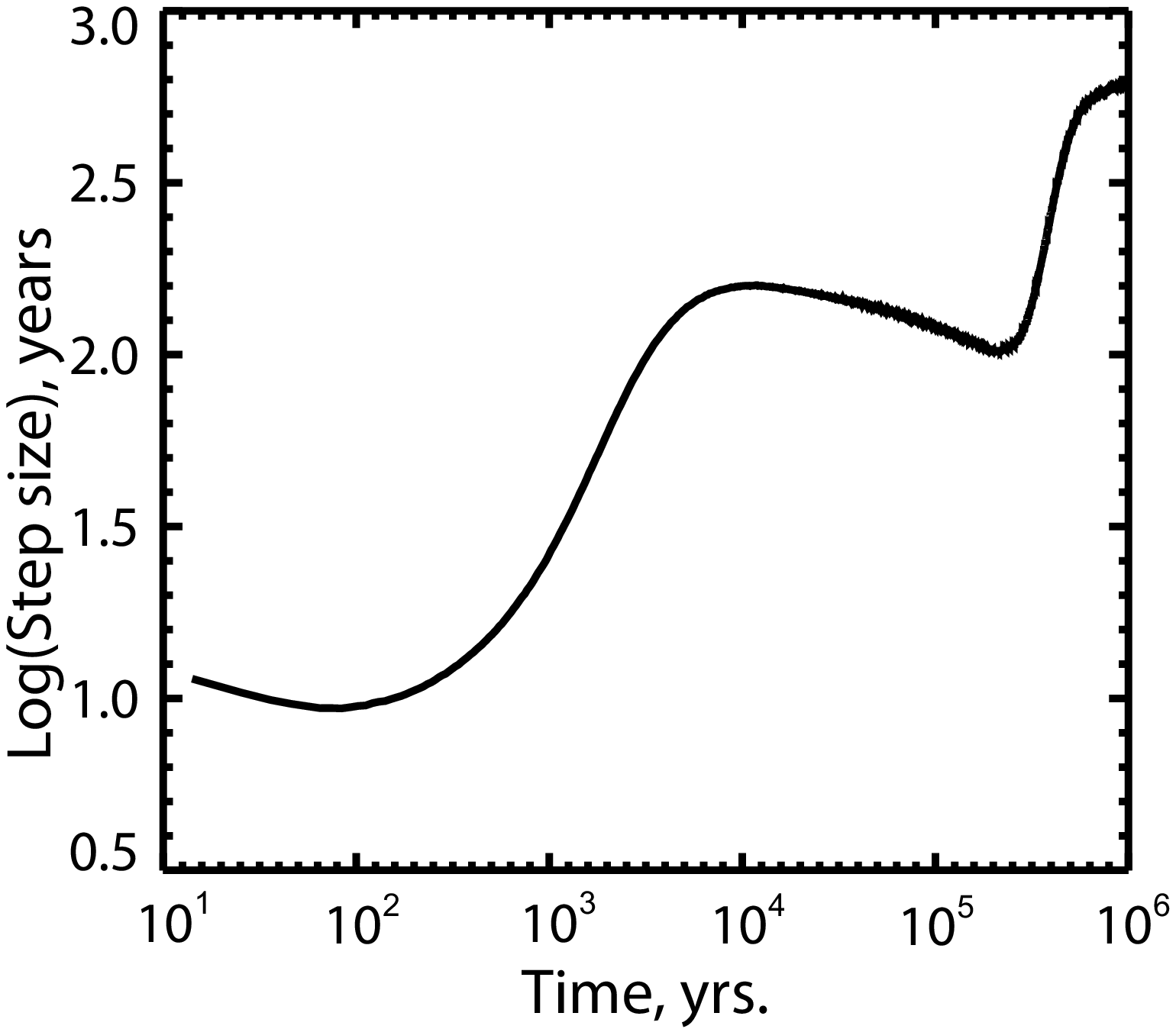} \caption{The Monte Carlo time step as a function of
time in the model of the chemical evolution of a dense cloud.}
\label{TimeStep_vs_Time}
\end{figure}

Due to the fact than the Monte Carlo technique operates with
integers, the smallest abundance of a molecule is 1. Given the dust
properties in our model (radius $r=10^{-5}$ cm, density $\rho_{\rm
dust}=3$ g cm$^{-3}$), a dust-to-gas ratio of 1/100 means that
volume per one grain is about $10^{12}/n({\rm H})$ cm$^3$, where
$n({\rm H})$ is the number density of hydrogen nuclei. Therefore, an
absolute population of 1 corresponds to a relative abundance of
$10^{-12}$ with respect to the total number of hydrogen atoms (see
Eq. \ref{gasvol_pergrain}). This is the lowest abundance directly
resolvable in our model. However, the huge amount of tiny time steps
taken during the calculations allows to average stochastic
abundances over wider time spans of 10--1\,000 years and push the
smallest abundance resolvable by the Monte Carlo code below
$10^{-12}$:
\begin{equation}
\overline{X}=\frac{\sum_{i=1}^{N}{X_{i}\Delta t_{i}}}{\sum_{i=1}^{N}{\Delta t_{i}}}
\end{equation}
where $X_{i}$ is the abundance of species after the $i$th time step,
the $\Delta t_{i}$ is the time step in seconds and $N$ is the amount
of time steps over which the averaging is performed. In such an
averaging procedure, the noise in the random variable $X_{i}$
decreases with $N$ as $\sqrt{N}$ and enables us to resolve
abundances below $10^{-12}$. In our simulations we averaged
abundances over $10^{6}$ time steps. Therefore, the resolution of
our calculations is about 10$^{-15} - 10^{-16}$ cm$^{-3}$. The
average value of the Monte Carlo time step in Gillespie's algorithm
is equal to an inverse sum of rates of all chemical reactions in the
model. For example, in the dense cloud model many molecules form
rapidly at initial times and less actively toward the end of
evolution (Fig.~\ref{TimeStep_vs_Time}). Time intervals,
corresponding to $10^{6}$ time steps in a medium with density
$10^{4}$ cm$^{-3}$ vary from 10 to $7\cdot10^{2}$ years for early
and late time moments, correspondingly.

\section{Comparison of the Monte Carlo method with rate equations}

While the Monte Carlo (MC) technique seems to be the most adequate
method to solve the chemical master equation, it is computationally
extremely demanding. Rate equations (RE) do not account for the
stochastic nature of surface processes and thus may produce spurious
results in some circumstances. On the other hand, they do require
much less computer power and are easier to handle. Therefore rate
equations will remain an important tool in theoretical
astrochemistry.

Given the importance of surface processes, it is thus necessary to
isolate regions in the parameter space where rate equations should
not be used at all or, at least, should be used with caution.
Similar comparisons, having been made so far, are based on a very
limited number of species and reactions \citep[e.g.][]{Charnley98,
Charnley01, RuffleHerbst00, Green_ea01, Caselli_ea02,
Stantcheva_ea02, StantchevaHerbst03, StantchevaHerbst04,
LipshtatBiham03, BarzelBiham07, Chang_ea07}. As we implemented a
unified Monte Carlo approach to the gas-phase and surface chemistry,
which is capable to treat a ``regular size'' chemical network, we
are able to perform a comprehensive comparison between predictions
of the MC and the (M)RE approaches.

For both the Model~T and
the Model~H, the rate equations (RE), the modified rate approach
(MRE) (for surface chemistry) and the Monte Carlo method are used to
simulate gas-phase and surface chemistry. The differences in
abundances discussed are related to surface chemistry and/or
dust-gas interactions. For the gas-phase chemistry, all models
produce identical results. Each run represents a combination of the tunneling treatment (T or H) and the surface chemistry treatment (RE, MRE, or MC). In total, we considered six models denoted as T-RE, T-MRE, etc.

\subsection{Global Agreement}

\begin{figure}
\plotone{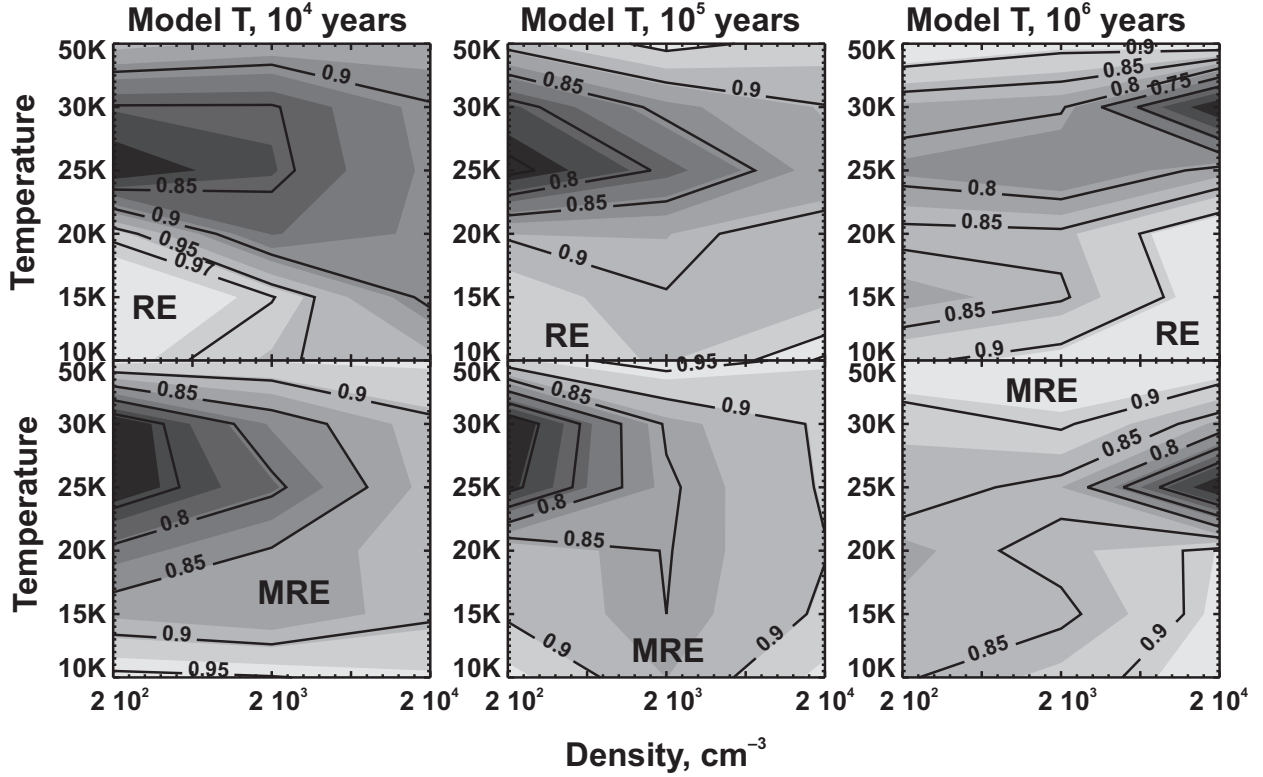} \caption{Global agreement diagrams for Model T at
three time moments. Percentage of species is shown in contour
labels, for which abundances computed with the rate equations and
with the Monte Carlo method differ by no more than an order of
magnitude. Grayscale map with arbitrary contours is added for
clarity. Darker color corresponds to worse agreement. Results are
presented for conventional rate equations (top row) and for modified
rate equations (bottom row).} \label{hasg_rate}
\end{figure}

\begin{figure}
\plotone{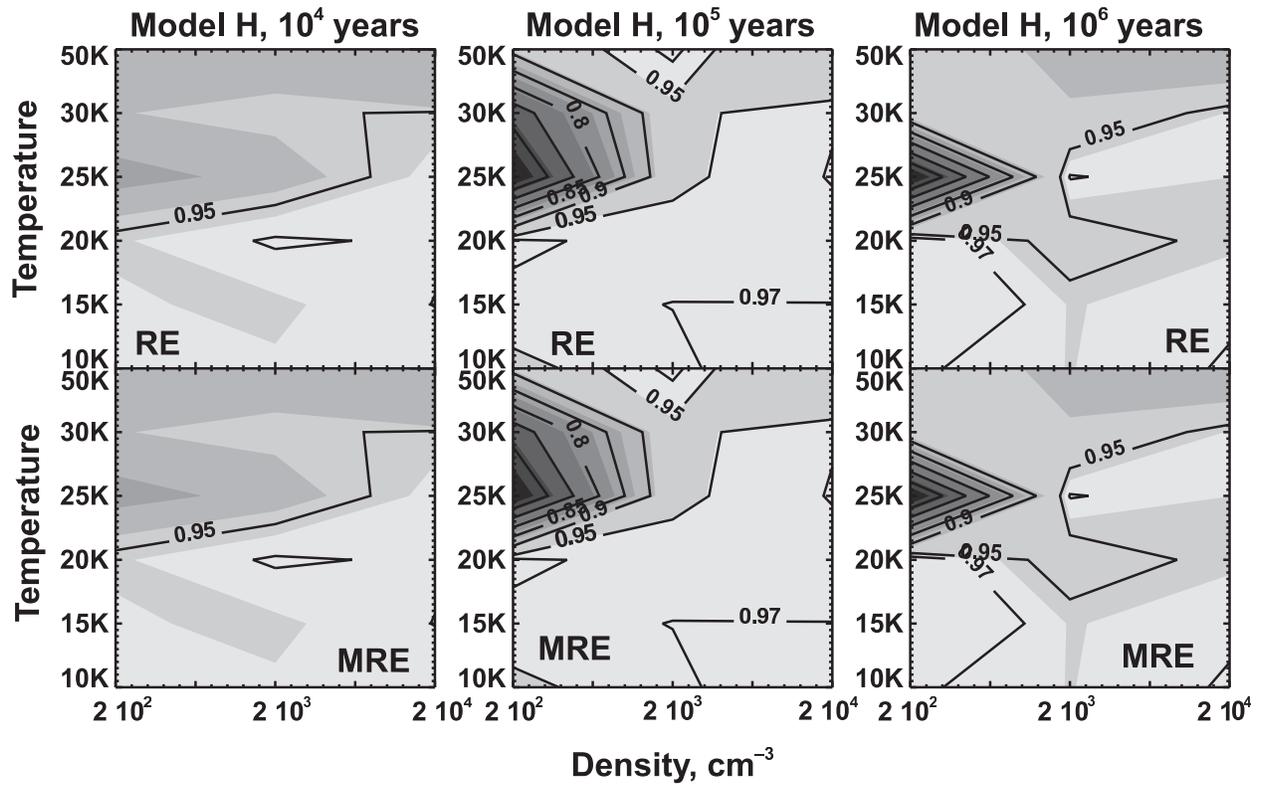} \caption{Same as in Figure~\ref{hasg_rate}, but for
Model H.} \label{obih_rate}
\end{figure}

To perform an overall comparison, in Figures~\ref{hasg_rate} and%
~\ref{obih_rate} we show diagrams of global agreement between MC,
RE, and MRE calculations as a function of $n$ and $T$ for
evolutionary times $10^4$, $10^5$, and $10^6$ years. For a given
species, the two methods, being compared, are assumed to agree at
time $t$ if they produce abundances, that differ by no more than an
order of magnitude. As we have mentioned above, in the MC method we
are able to resolve abundances down to $10^{-15}$, which would lead
to a formal disagreement between abundances, say, $10^{-15}$ in the
MC method and $10^{-16}$ (or smaller) in the RE method (all abundances are
relative to the total number of H atoms). This kind of
disagreement is not meaningful from the observational point of view
because such small relative abundances are extremely difficult to
observe with a satisfactory accuracy. In addition, even though lower
abundances can be calculated by the RE method, their actual accuracy
is limited by the numerical interpolation. In the following
discussion, species, for which both methods predict relative
abundances less than $10^{-12}$, are excluded from further
consideration. In Figures~\ref{hasg_rate} and ~\ref{obih_rate},
contours labeled with 0.9 mean that the two methods give abundances
that differ by less than an order of magnitude for 90\% of species
with abundances higher than $10^{-12}$ in both methods.

\begin{figure}
\plotone{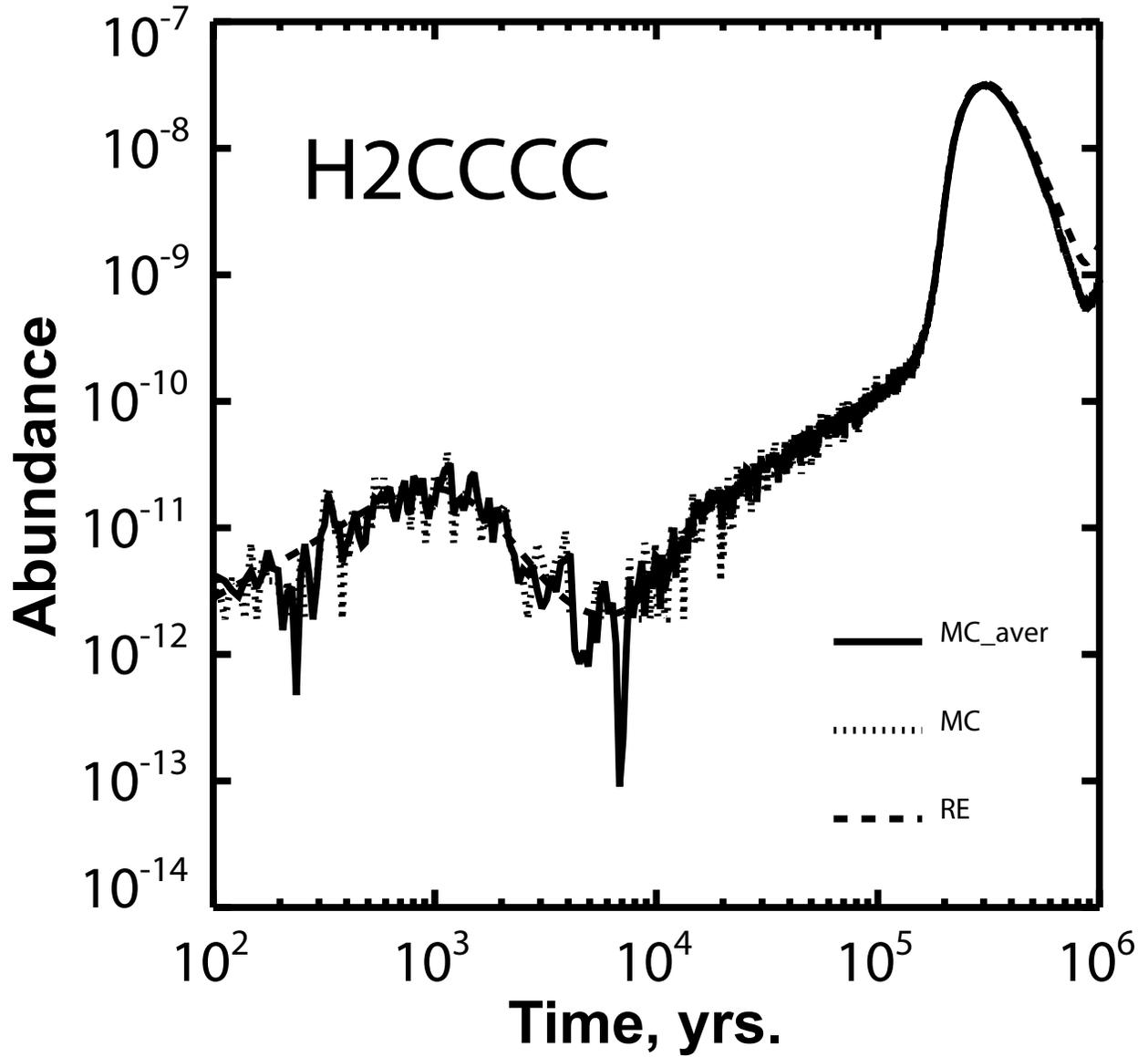} \caption{Evolution of H$_2$CCCC gas-phase abundance at
$n=2\cdot10^4$~cm$^{-3}$ and $T=10$\,K, simulated with the MC and RE
methods (Model~T). Dotted line indicates raw MC results, solid line
shows MC results after averaging. Results of RE integration are
shown with dashed line.} \label{h2cccc}
\end{figure}

An order-of-magnitude agreement criterion may seem to be too coarse.
However, when stochastic effects do not play an important role, all
three method produce results, which are nearly identical. To
illustrate this, we show in Figure~\ref{h2cccc} the evolution of the
H$_2$CCCC abundance at $n=2\cdot10^4$~cm$^{-3}$ and $T=10$\,K,
simulated with the MC and RE methods. The species is chosen because
its abundance strongly varies with time. It can be seen that the
difference between the two methods is very small.

\subsubsection{Model~T}

In Model~T the species are highly mobile---atomic and molecular
hydrogen because of tunneling, and all the other species because of
the low energy barrier for diffusion. Mobility drives rapid surface
reactions with rates which sometimes exceed the accretion/desorption
rate. Surface reactions, which are on average faster than accretion,
are not permitted in the MC approach when surface populations of
their reagents are low ($\sim$~1), but they are allowed in the RE
approach, where the surface abundance of a species can be much less
than one. Because of that, differences between Monte Carlo and rate
equation methods are quite noticeable in Model~T.

At temperatures below 20\,K and above $\sim40$\,K the agreement
between the stochastic method and the rate equation method (both RE
and MRE) is about 85\% or better at all times and densities
(Figure~\ref{hasg_rate}). Around $T\sim10$\,K the residual
discrepancy is mainly caused by complex surface species, which have
zero abundances in Model~T-MC and are overproduced in Model~T-RE and
Model~T-MRE, so that their abundances are just above the adopted
cutoff of $10^{-12}$. If we would raise the cutoff to $10^{-11}$,
the agreement would be almost 100\%. At $T\sim50$\,K the agreement
is also nearly perfect, as the chemistry is almost a purely
gas-phase chemistry under these conditions (dust temperature being
equal to gas temperature).

All the major discrepancies are concentrated in the temperature
range between 25--30\,K. In this range the accretion rate (which
depends on gas temperature) is high enough to allow accumulation and
some processing of surface species, while the correspondingly high
desorption rate (which depends on dust temperature and $A_{\rm V}$)
precludes surface production of complex molecules. The latter
process is adequately described in the MC runs only.

At earlier times ($t\la10^5$ years) the largest differences in this
temperature range are observed at lower densities. They are caused by
an overproduction of `terminal' surface species, like water,
ammonia, hydrogen peroxide, carbon dioxide, etc., in the RE and MRE
calculations. At later times ($t\sim10^6$~years) situation changes
drastically. The discrepancies shift toward higher density, and the
overall agreement falls off to about 60\% for both RE and MRE
calculations. Even though the stochastic chemistry is only important
on dust surfaces, by this time it also influences many gas-phase
abundances due to effective gas-grain interactions.

At $t=10^6$~years, in the RE calculation at $T=30$\,K, among 234 species with
abundances higher than $10^{-12}$, 95 species (including 81
gas-phase species) have an order of magnitude disagreement with the
MC run. In the MRE run at $T=25$\,K, the number of species with
abundances above $10^{-12}$ is 221. Among them 83 species disagree
with the MC results by more than an order of magnitude. In both
calculations these are mostly carbon-bearing species, like CO,
CO$_2$, CS, to name a few, and, in particular, carbon chains (like
some cyanopolyynes). It is noteworthy that not only the abundances
of neutral species disagree, but also the abundances of some ions,
including C$^+$ and S$^+$.

\subsubsection{Model~H}

In Model~H the mobility of all species on the surface (including H
and H$_2$) is caused by thermal hopping only, which is slow because
of the high $E_{\rm b}/E_{\rm D}$ ratio. One of the consequences of
these slow surface reaction rates is that results produced by the
modified rate equations almost do not differ from results of
conventional rate equations. The essence of the modification is to
artificially slow down surface reactions to make their rates
consistent with accretion/desorption rates. In Model~H all surface
reactions are slow anyway, so that modifications never occur. Thus,
we only compare the results of the MC and RE calculations.

At $t=10^4$ years the percentage agreement between the two runs
never falls below 80\% and most often is actually much better than
the adopted order-of-magnitude criterion.
At later times, the discrepancy appears in the same temperature
range as in Model~T, i.e., between 25 and 30\,K. Unlike Model~T,
both at $t=10^5$~years and $t=10^6$~years the disagreement shows up
at low density. The set of discrepant species is nearly identical at
both times and consist mostly of ices, overproduced by rate
equations. These ices include some key observed species. For
example, NH$_{3}$, H$_{2}$O, CO$_2$ ices are all abundant in the
Model~H-RE despite the low density, with a surface H$_{2}$O
abundance reaching $3\cdot10^{-7}$ by $10^6$ years. On the other
hand, the Model~H-MC results in very small or zero abundances for
the same species. More chemically rich ices in Model~H-RE bring about
enhanced gas-phase abundances for the same molecules. The only
species which is {\em under}produced by the rate equation approach
is surface C$_2$O.

\subsection{Selected Species}

The  diagrams presented above indicate areas in the parameter space,
relevant to molecular clouds, where deterministic methods fail to
describe stochastic surface processes. However, a global agreement
does not necessarily imply that the abundances of key species are
also correctly calculated. In the following we discuss a few
important gas-phase species and ices, for which an order of
magnitude (or more) disagreement has been found.

\subsubsection{Model~T}

\begin{figure}
\plotone{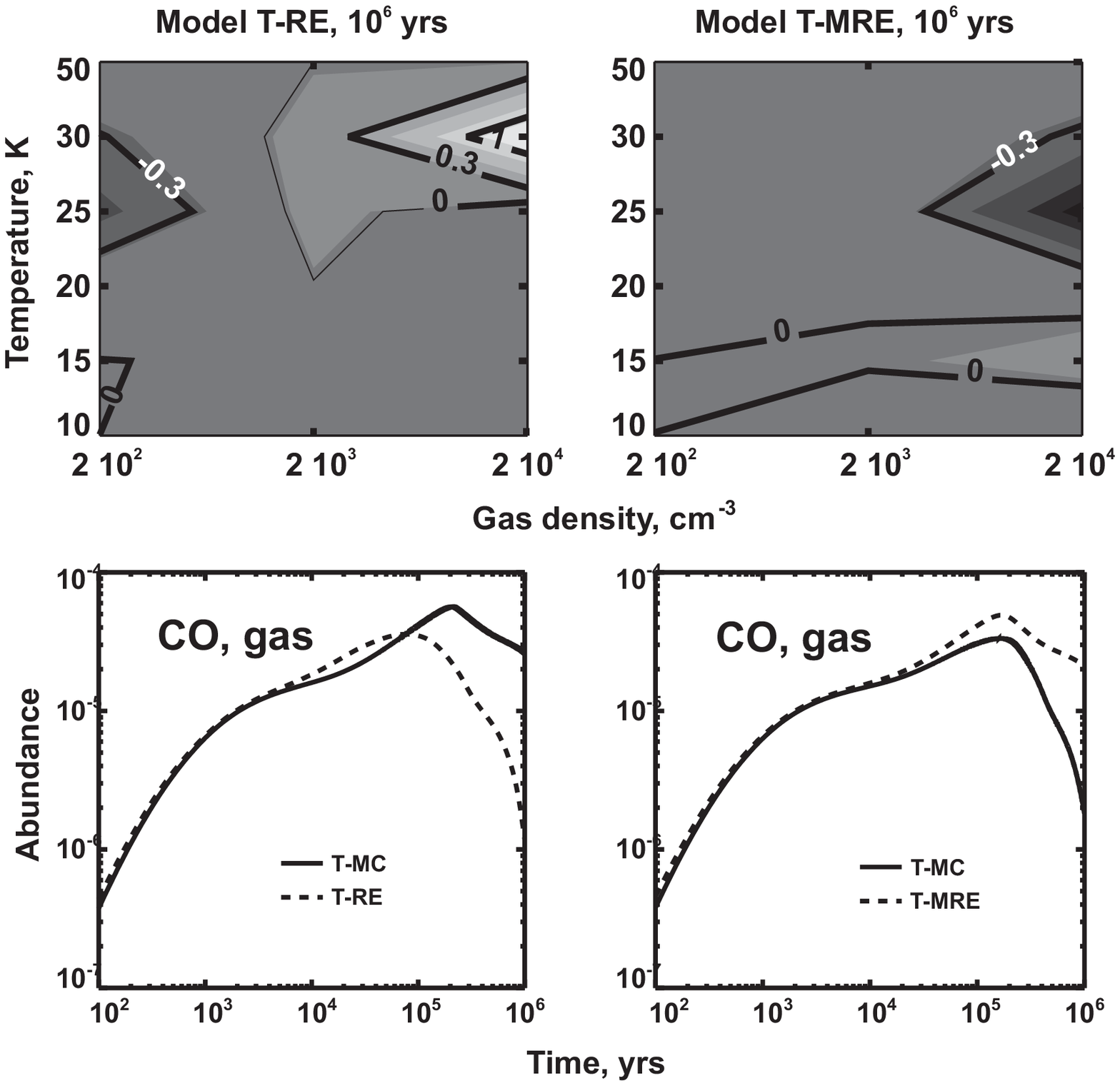} \caption{(Top row) Agreement diagrams for CO in Model~T.
Contour labels correspond to the logarithm of the ratio of gas-phase CO
abundance in the MC run to that in the RE (top left) and MRE (top
right) runs. (Bottom row) Gas-phase CO abundance evolution in the RE run
($n=2\cdot10^4$~cm$^{-3}$ and $T=30$\,K; bottom left) and in the MRE
run ($n=2\cdot10^4$~cm$^{-3}$ and $T=25$\,K; bottom right).}
\label{COtwo}
\end{figure}

Because surface reaction rates are high in this model, there are
many species for which stochastic and deterministic methods give
quite different results. The most abundant trace molecule, CO, is
among these species. In Figure~\ref{COtwo}, we show agreement
diagrams for Models T-RE and T-MRE (top row) and the CO abundance
evolution for $n-T$ combinations corresponding to the worst
agreement. Both the RE and MRE methods fail to reproduce the late CO
evolution at high density and temperatures of 25--30\,K (where the
global agreement is also worst in this model), but each in a
different way. While in Model~T-RE CO is underabundant by an order
of magnitude, in Model~T-MRE it is overabundant by the similar
amount. This difference is related to the treatment of CO
$\longrightarrow$ CO$_2$ conversion on dust surfaces, where
underabundance of the CO ice in Model~T-RE and overabundance in
Model~T-MRE are observed at almost all times. At later time, these
surface abundances just start to propagate to gas-phase abundances.

Overproduction of the CO$_2$ ice in Model~T-RE also consumes carbon
atoms which would otherwise be available not only for CO, but also
for more complex molecules, in particular, carbon chains, starting
from C$_2$, observed in diffuse clouds, and C$_2$S, used as a
diagnostics in prestellar cores. In the RE and MRE calculations
these molecules, like CO, show trends opposite to that of CO$_2$
ice: when surface CO$_2$ is overabundant with respect to the MC
model, carbon chains are underabundant and vice versa.

When CO molecule sticks to a grain, it either desorbs back to the gas-phase,
where it may participate in further processing, or is converted into CO$_2$.
As the desorption energy is quite high for CO$_2$, it acts as a sink
for carbon atoms at these temperatures. In the accretion-limited regime,
rate equations overproduce CO$_2$ with respect to the MC method. Thus,
we see less CO and other carbon-bearing species in the gas-phase. The MRE
method helps, probably, too much to account for accretion-limited CO
processing and essentially quenches surface CO $\longrightarrow$ CO$_2$
 conversion. It is worth noting that at high density in this particular temperature range
(and for adopted CO desorption energy) CO balances between complete
freeze-out (at $T < 20$\,K) and near absence on dust surface (at $T>30$\,K).
Thus, even relatively minor changes in treatment may lead to noticeable consequences.

\begin{figure}
\plotone{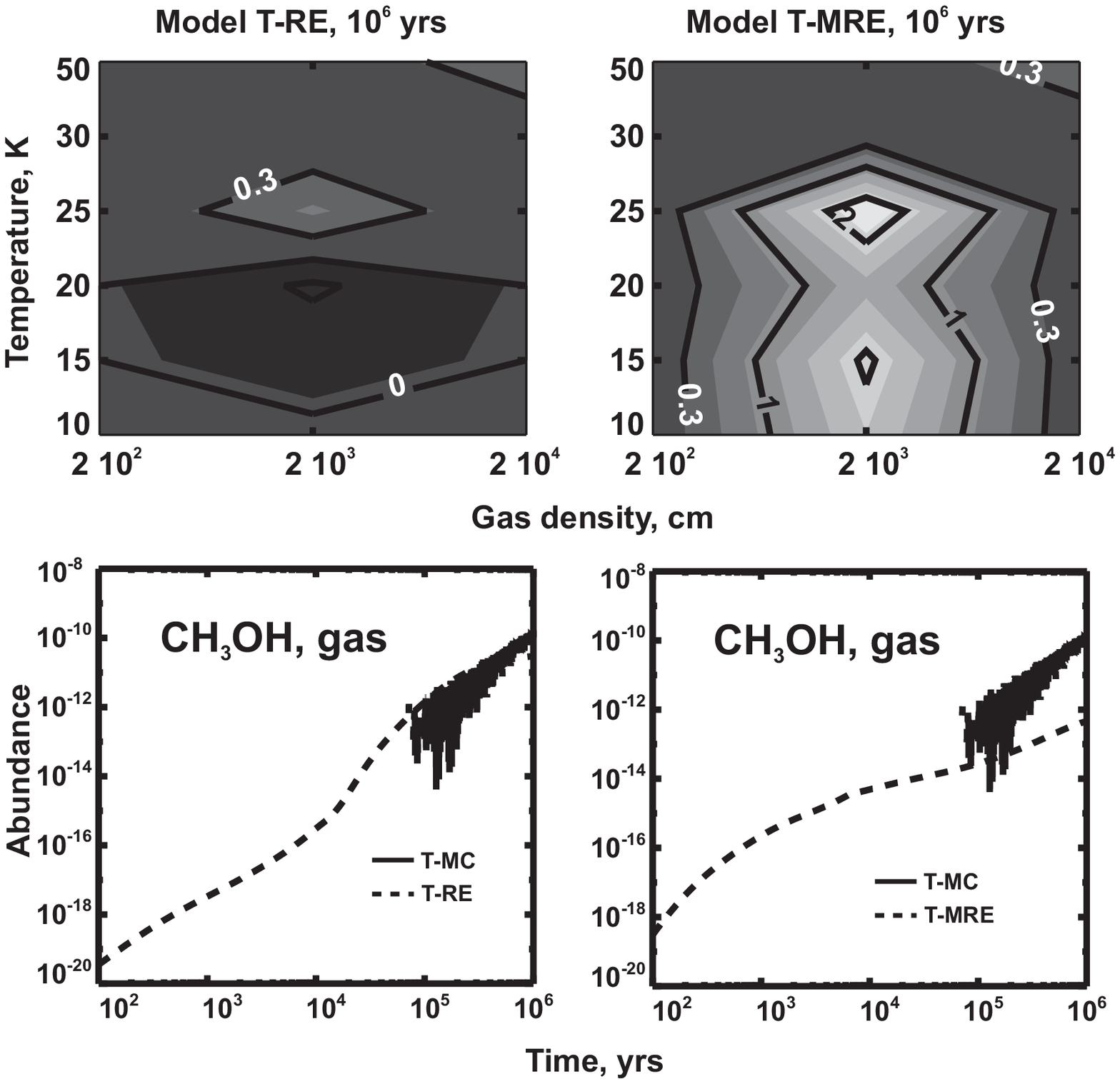} \caption{(Top row) Agreement diagrams for methanol.
Contour labels correspond to the logarithm of the ratio of gas-phase CH$_3$OH
abundance in the MC run to that in the RE (top left) and MRE (top
right) runs. (Bottom row) Gas-phase methanol abundance evolution in the RE run
($n=2\cdot10^3$~cm$^{-3}$ and $T=25$\,K; bottom left) and in the MRE
run ($n=2\cdot10^3$~cm$^{-3}$ and $T=25$\,K; bottom right).}
\label{CH3OHtwo}
\end{figure}

Another carbon-bearing molecule that shows an interesting behavior
is methanol in the gas phase. Diagrams, comparing its abundance
computed with the MC approach and with the rate equations, are
presented in Figure~\ref{CH3OHtwo}. The agreement between results of
the MC calculations and the RE calculations is almost perfect, at
least, within the scatter produced by the MC simulations. However,
the modified rate equations, which tend to improve agreement between
stochastic and deterministic methods for many other species, in this
particular case underpredict methanol abundance by more than two
orders of magnitude in comparison with the MC run. The same is true
for surface methanol as well. It is interesting to note that the region in the
parameter space, where MC and MRE methanol abundances disagree, extends
down to lowest temperatures considered.

\begin{figure}
\plotone{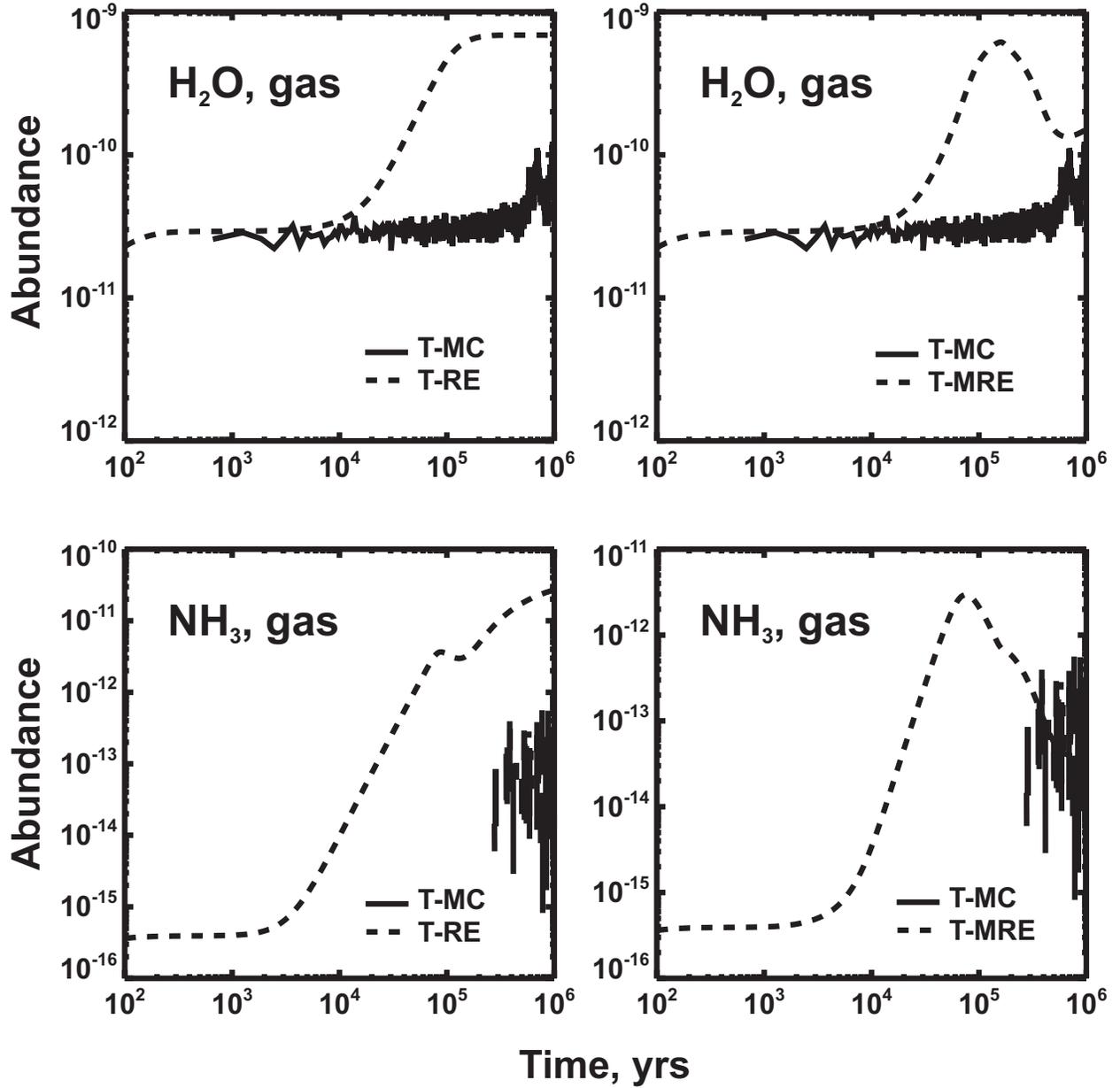} \caption{Water and ammonia gas-phase abundance evolution in
the RE run in Model~T ($n=2\cdot10^2$~cm$^{-3}$ and $T=25$\,K; left) and in the
MRE run ($n=2\cdot10^2$~cm$^{-3}$ and $T=25$\,K; right).}
\label{water}
\end{figure}

An interesting pattern is presented by the abundance of water,
ammonia and their ices (Figure~\ref{water}). These species are
severely overproduced by the RE methods at $t=10^5$~years, but after
this time the modified rate equations are able to restore the
agreement with the MC method, so that by $10^6$ years difference
between MC and MRE runs is not significant. This is an example of a
situation where the MRE help to produce realistic results, at least,
at later times.
We show plots only for gas-phase abundances
because these are what is really observed. However, these abundances
are only a reflection of surface processes, and surface water and
ammonia abundances behave in exactly the same way. It is
hard to name a single reaction which is responsible for the
difference between results of RE and MRE methods in this case.
Let's consider water as its chemistry is somewhat simpler.
The primary formation reaction for water is H~+~OH, with
only a minor contribution from H$_2$O$_2$ (see below).
However, hydroxyl is not produced in H~+~O reaction, as one
might have expected. Because of the paucity of H atoms
on the grain surface, hydroxyl formation is dominated by reaction
\[
{\rm O}+{\rm HNO}\longrightarrow{\rm NO}+{\rm OH}.
\]
Abundance of HNO is restored in reaction
\[
{\rm H}+{\rm NO}\longrightarrow {\rm HNO}.
\]
These two reactions form a semi-closed loop for which
the sole result is OH synthesis out of an H atom and an O atom.
Evolution of surface O abundance seems to be the key to the
difference between the RE and MRE models. In the RE model
the number of O atoms on a surface is determined by the balance
between their accretion from the gas phase and consumption
in O~+~HNO reaction. The same processes define O abundance
in the MRE model as well, but only at $t\la10^5$ years. After this
time O+HNO reaction slows down significantly due to modifying
correction, and O abundance is controlled purely by accretion and
desorption thereafter (as it is in the MC model at all times). As
a result, OH and H$_2$O abundances decrease, getting closer to the
MC model prediction. A similar mechanism is at work in the case
of ammonia as well.

\subsubsection{Model~H}

The only parameter region where a discrepancy occurs for Model~H is
located at $n=2\cdot10^2$~cm$^{-3}$ and $T=25$\,K. Because of the
low density, the degree of molecular complexity at these conditions
is also low, and by $10^6$ years only 73 species have abundances
higher than $10^{-12}$ at least in one of the calculations. Of these
73 species, 25 show disagreement between the MC calculations and the
RE calculations.

\begin{figure}
\plotone{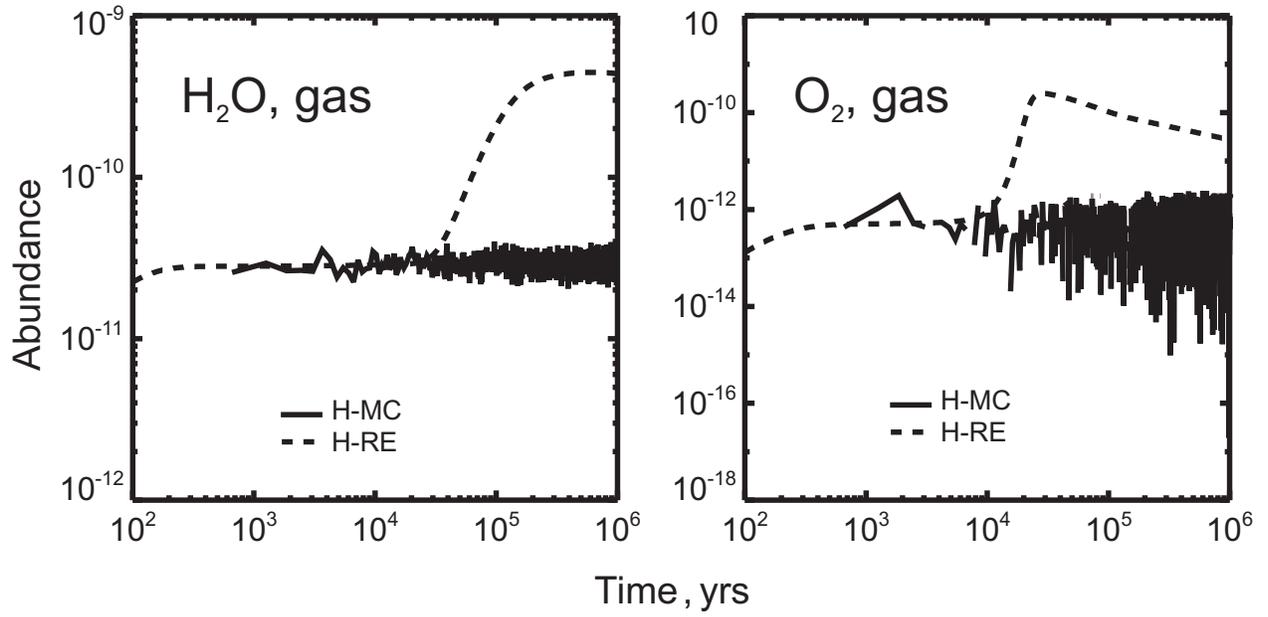}
\caption{Water and molecular oxygen abundances at $n=2\cdot10^2$~cm$^{-3}$
and $T=25$\,K in Model~H.}
\label{model_h}
\end{figure}

In the top row of Figure~\ref{model_h} we show the low density
evolution of gas-phase water and molecular oxygen abundances. Both
plots show a similar trend. In the MC calculations the abundance of
the molecule stays nearly constant. At earlier times this
steady-state behavior is reproduced in the RE run as well, but later
the abundance grows and exceeds the MC abundance by more than an
order of magnitude. It is interesting to note that the {\em
average\/} O$_2$ abundance in the MC run seems to decrease somewhat
at later times, and this trend is reproduced by the RE model. The behavior
of gas-phase abundances is a reflection of grain-surface abundances.
It must be kept in mind that surface water abundance is quite high in
the H-RE model, reaching almost $10^{-6}$. Even though evaporation rate
is not very high, it does increase gas-phase abundance up to a level of a
few time $10^{-10}$. Also, it is a low density model, so desorption is
primarily caused by photons.

\begin{figure}
\plotone{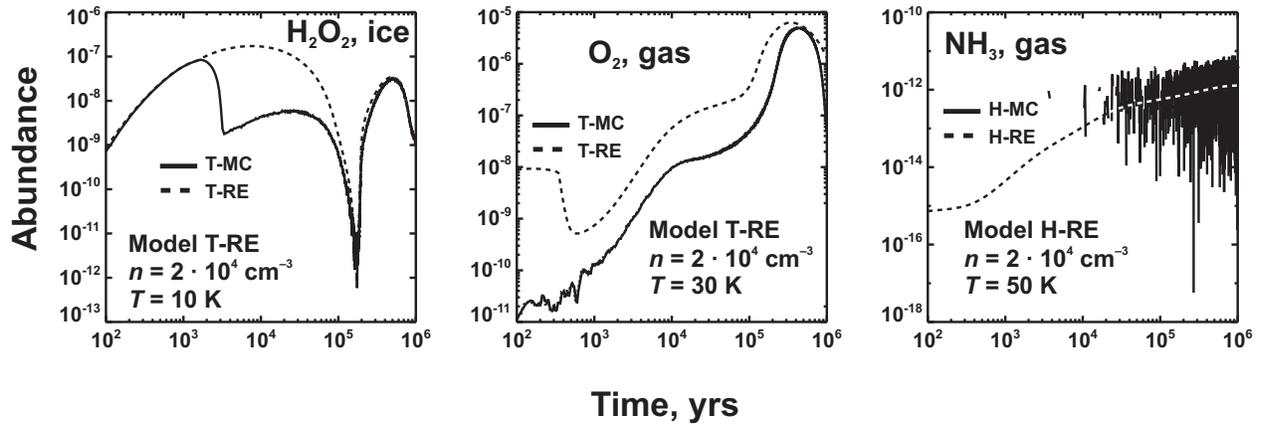} \caption{Abundance evolution for selected species.
Models and physical conditions are indicated on the diagrams.}
\label{misc}
\end{figure}

In Figure~\ref{misc} we show the evolution of H$_2$O$_2$ surface
abundance and O$_2$ gas-phase abundance in Model~T-RE at
$n=2\cdot10^4$~cm$^{-3}$ and the gas-phase ammonia abundance in
Model~H-RE at $n=2\cdot10^3$~cm$^{-3}$. These plots demonstrate that
disagreement does not always mean simply higher or lower abundances.
Different processes determine the H$_2$O$_2$ abundance at different
times, and while the RE method is able to capture some of these
processes, others are obviously missed. Our analysis shows that this behavior
is related to the treatment of reaction with H atoms. Specifically, surface
abundance of H$_2$O$_2$ is controlled by three reactions only.
It is produced in a reaction H + O$_2$H and destroyed by atomic hydrogen
in reactions producing either H$_2$O + OH or O$+2$H + H$_2$. (Note
that in the considered model water ice is almost irreversible sink
for O atoms as there are no surface reactions destroying water, and
desorption is negligible.) The formation reaction is fast as it does not have an
activation barrier. So, initially, as H and O atoms start to stick to a grain,
surface abundance of H$_2$O$_2$ steadily grows. At the same time,
there is not enough H atoms for either of destruction reactions. Then, in the MC model
at around 1300 years the first destruction reaction (H~+~H$_2$O$_2$
$\longrightarrow$ H$_2$O~+~OH), which has a slightly lower activation
barrier than the other one (H~+~H$_2$O$_2$ $\longrightarrow$ O$_2$H~+~H$_2$),
starts to ``steal'' some hydrogen atoms from the formation reaction, irreversibly
removing one O atom per reaction from the H$_2$O$_2$ (re)formation process.
This is the reason for the sharp fall-off of H$_2$O$_2$ abundance. On the
other hand, in the RE model both destruction and formation reactions may
occur simultaneously, which prevents the first destruction reaction from
having such a dramatic effect. Note that the second reaction restores
an O$_2$H molecule which may react again with an H atom to restore an
H$_2$O$_2$ molecule.

H$_2$O$_2$ is not unique in this kind of behavior. There are some other
surface species which show more or less similar trends, that is, almost
perfect agreement at some times and quite noticeable disagreement
at other times. These are some carbon chains (C$_2$H$_2$, C$_2$N)
or simpler molecules (CS, NO). Another equally dramatic example
is represented by C$_2$H$_2$. The mechanism is similar, that is,
it is related to the sequence of hydrogen additions which is treated
differently in the MC model and the RE model.

Evolution of the O$_2$
abundance computed with the RE method follows the evolution computed
with the MC model, but most of the time the RE curve is somewhat
higher than the MC curve. Finally, the plot for ammonia shows the
situation when the RE method correctly describes the average
abundance evolution. However, the abundance predicted by the MC
method fluctuates so wildly, that at each particular $t$ we can
detect disagreement with high probability, using our formal
order-of-magnitude criterion.

\section{Discussion}\label{diss}

In the previous sections we investigated the validity of the
(modified) rate equation approach to grain surface chemistry under
different physical conditions  encompassing diffuse clouds, giant
molecular clouds, infrared dark clouds etc. Evolution of the medium
is studied with an extended gas-grain chemical model over a time
period of 10$^{6}$ years. We utilize the unified stochastic Monte
Carlo approach, applied simultaneously to gas-phase and
grain-surface reactions. The results are used to test the validity
of conventional deterministic approaches. In general, differences in
results obtained with deterministic and stochastic methods strongly
depend on the adopted microphysical model of surface chemistry. In
Model T, where tunneling for atomic and molecular hydrogen is
permitted and diffusion/desorption energies ratio is low ($E_{\rm
b}/E_{\rm D}=0.3$), discrepancy both for RE and MRE methods is very
significant. At low temperatures (10~-~20\,K) only abundances of
surface species are discrepant by more than an order of magnitude.
At moderate temperatures (25--40\,K), due to active gas-grain
interaction, incorrect treatment of grain surface chemistry becomes
important for gas-phase abundances, too, and leads to dramatic
decrease of overall agreement. At even higher temperatures agreement
becomes better again due to limited surface chemistry. In Model H
with no tunneling and high $E_{\rm b}/E_{\rm D}$ ratio, surface
species are much less mobile, and the agreement between
deterministic and stochastic methods is better. The only exception
is low density ($n({\rm H})=10^{2}$~cm$^{-3}$) with moderate
temperatures (25~--~30\,K), where average residence time of species
on grain surface is still lower than the average interval between
reaction events. In general, the stochastic chemistry severely
affects abundances and, thus, must be taken into account in chemical
models of a warm and moderately dense medium.

Modifications in rate equations, aimed at taking into account the
accretion-limited regime, do not provide a significant improvement
over `canonical' rate equations in Model~T. Close inspection of
Figure~\ref{hasg_rate} shows that at earlier times
($t\le10^5$~years) modified rate equations mostly produce results,
which are rather {\em less\/} consistent with the `exact' Monte
Carlo solution, than results of the RE method. Only at later times
($t\sim10^6$~years) results of the MRE method become more consistent
with the `exact' solution than the RE results. In the previous
section, we have already shown some examples of species, for which
the MRE method actually worsens the agreement.

Because at many of considered physical conditions results obtained
with Monte Carlo method deviate significantly from those obtained
with (modified) rate equations, the natural question to ask is
whether the inability of the RE method to treat these combinations
of parameters might have caused mis-interpretation of observational
data. To answer this question, one would need to consider an object,
for which a large volume of observational data is available, so that
simultaneous comparison for many molecules is possible. So far,
there seems to be just one molecular cloud, which is studied in
necessary details. This is a sub-region of the TMC1 cloud where
cyanopolyynes reach high concentrations, the so-called TMC1-CP.
The physical conditions in this object ($T \sim 10$~K, $n_{\rm H}
\sim 10^4$~cm$^{-3}$) do fall into the range adopted in our model
\citep{BM_89}. However, the TMC1-CP is probably too cold to show
significant dependence of gas-phase abundances from surface
processes.

We may expect that the disagreement will be more significant
in hotter objects. The range of physical conditions studied in this
paper is quite narrow, as it has only been chosen to provide an
initial view on the importance of stochastic treatment for surface
chemistry. There are two directions which are worth to follow for
further study. One of them is related to diffuse clouds. In this
paper we most of the time assumed initial conditions in which all
the hydrogen is molecular even at low density. Another
simplification, which is, strictly speaking, not valid in low
density medium, is equality of dust and gas temperature. So, it is
interesting to check if stochastic effects work in predominantly
atomic gas with different gas and dust temperatures. Another
direction is chemistry at later stages of star formation, that is,
in hot cores, hyper- and ultracompact HII regions, protoplanetary
disks. In particular, in dark cold disk midplanes, which are poor in
atomic hydrogen, stochastic effects may severely influence surface
synthesis of hydrogenated species, like formaldehyde, which are
later transferred by turbulent mixing in warmer disk areas and
desorbed into gas phase \citep{aikawa_co_disk}. In these
circumstances one would definitely need to use more accurate methods
to model surface chemistry. Stochastic effects may be even more
pronounced in the dynamically evolving medium, like during slow
warm-up phase in hot cores or ``corinos''. Complex (organic)
molecules will be among the most affected species.

While the Monte Carlo method is the most direct approach, it is by
no means practical. So, other methods are to be developed, in
particular, the method of moment equations, if we want to understand
deeper the chemical evolution in warm and dense astrophysical objects.

\section{Conclusions}

In this study, a gas-grain chemical model is presented, consisting
of about 600 species and 6000 reactions with stochastic description
of grain surface chemistry. For the first time both gas phase and
grain surface reactions are simulated with a unified Monte Carlo
approach. This unified model is used to test the validity of rate
equations and modified rate equations
for a set of physical conditions, relevant to translucent
clouds, dark dense cores, and infrared dark clouds.

A comparison of results obtained with deterministic rate equations
(RE) and modified rate equations (MRE) with results of Monte Carlo
simulations was performed for this set of physical conditions and
two different models of surface chemistry. We found
that results obtained with both RE and MRE approaches sometimes significantly
deviate from those obtained with Monte Carlo calculations in Model~T, 
where surface species are highly mobile and most of the grain
chemistry occurs in the accretion limited regime. While at low
temperatures (10\,K -- 20\,K) mainly RE and MRE abundances of
surface species show significant discrepancies, at moderately high
temperatures (25\,K -- 30\,K) even abundant gas-phase species like
CO considerably deviate from the Monte Carlo results. At these
temperatures grain surface chemistry is still very active and at the
same time processes of accretion and desorption are very efficient.
The gas-phase abundances of many species are influenced by surface
chemistry that is correctly described in Monte Carlo simulations
only. In Model H surface species are much less mobile than in Model
T and the agreement between the (modified) rate equations approach
and Monte Carlo simulations is almost perfect. The only parameter
region where this discrepancy is significant is at a temperature of
$T=25$\,K and at a low density $n({\rm H})=2\cdot10^{2}$ cm$^{-3}$.
On the whole, results of modified rate equations seem to be in
somewhat closer agreement with Monte Carlo simulations than results
of the conventional rate approach. But still, the agreement is far
from being perfect, so modified rates should be used with care. We
conclude that stochastic effects have a significant impact on
chemical evolution of moderately warm medium and must be properly taken
into account in corresponding theoretical models. In the studied parameter range,
stochastic effects are, in general, not important at low ($T\sim10$\,K)
and high ($T\sim50$\,K) temperatures.

\acknowledgements
      Part of this work was supported by the Russian
      \emph{Dynasty} foundation. DW is supported by the RFBR grant 07-02-01031. Authors are grateful to Sergey Koposov for
valuable and fruitful discussions and assistance in data analysis
software development. Authors are
thankful to the anonymous referee for valuable comments and
suggestions. This research has made use of NASA's
Astrophysics Data System.


\clearpage
\begin{table}
\caption{Adopted surface chemistry models} 
\label{tab:surf_mods}      
\centering                          
\begin{tabular}{cccc}        
\hline\hline                 
 Model & Based on & Tunneling & $E_{\rm b}/E_{\rm D}$ \\    
 \hline
T & \citet{HHL92} & Yes & 0.3 \\
H & \citet{Katz_ea99} & No & 0.77 \\
\hline                        
\hline                                   
\end{tabular}
\end{table}

\clearpage

\clearpage

\clearpage

\clearpage

\clearpage

\clearpage

\clearpage

\clearpage

\clearpage

\end{document}